\documentclass[fleqn,usenatbib]{mnras}

\usepackage{newtxtext,newtxmath}
\usepackage{textcomp}
\usepackage{amsmath}
\usepackage{amsfonts}

\usepackage[T1]{fontenc}

\DeclareRobustCommand{\VAN}[3]{#2}
\let\VANthebibliography\thebibliography
\def\thebibliography{\DeclareRobustCommand{\VAN}[3]{##3}\VANthebibliography}

\usepackage{soul}

\usepackage{graphicx}	
\usepackage{amsmath}	
\RequirePackage{tikz}

\graphicspath{{images/}}

\title[GRG--ICM interaction]{Probing AGN duty cycle and cluster-driven morphology in a giant episodic radio galaxy}

\author[Kumari et al.]{
Shobha Kumari,$^{1}$\orcidA{}\thanks{E-mail: shobhakumari@mcconline.org.in}
Sabyasachi Pal,$^{1}$\orcidB{}
Surajit Paul$^{2}$\orcidC
and 
Marek Jamrozy$^{3}$\orcidD
\\
$^{1}$Midnapore City College, Kuturia, Bhadutala, Paschim Medinipur, West Bengal, 721129, India\\
$^{2}$Manipal Centre for Natural Sciences, Manipal Academy of Higher Education, Karnataka, Manipal, 576104, India \\
$^{3}$Obserwatorium Astronomiczne, Uniwersytet Jagielloński, ul. Orla 171, 30-244 Kraków, Poland\\ 
}

\newcommand{\orcidicon}{\includegraphics[width=0.32cm]{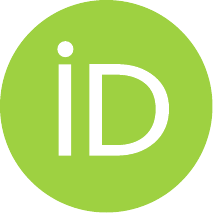}}

\foreach \x in {A, ..., Z}{%
\expandafter\xdef\csname orcid\x\endcsname{\noexpand\href{https://orcid.org/\csname orcidauthor\x\endcsname}{\noexpand\orcidicon}}
}

\begin{document}
\label{firstpage}
\pagerange{\pageref{firstpage}--\pageref{lastpage}}
\maketitle
\def \src{J1007+3540}

\begin{abstract}
The evolution of radio jet morphology and its energetics is significantly influenced by the environment in which the host galaxy resides. As giant radio galaxies (GRGs) often extend to the scale of entire galaxy clusters ($\sim$Mpc) and beyond, they are a suitable class of objects for studying jet--intracluster medium interactions. This paper presents a multiwavelength study of a GRG, J1007+3540, using LOFAR Two-metre Sky Survey second data release (LoTSS DR2) at 144 MHz and the upgraded Giant Metrewave Radio Telescope (uGMRT) at 400 MHz. The source has a projected linear extension of 1.45 Mpc and is hosted by MaxBCG J151.77665+35.67813, within the WHL 100706.4+354041 cluster. At both frequencies, the source exhibits clear signatures of recurrent jet activity, a one-sided, extended, tail-like diffuse structure with a morphological break in the tail. The estimated radiative ages of the inner lobes and outer north lobe are $\sim$140 Myr and $\sim$240 Myr, respectively. In addition to the radio analysis, we performed optical--to--infrared spectral energy distribution modelling. The host galaxy is an evolved elliptical system with a stellar mass of $\log_{10}(M_\star/M_\odot) = 11.0$ and an old stellar population age of $\sim$12 Gyr. The high infrared-derived star formation rate ($\sim106~M_\odot$~yr$^{-1}$) of the source implies significant dust-obscured star formation, potentially linked to merger-driven gas inflows. J1007+3540 presents a rare combination of a restarted jet, a detached tail-like structure, and unusual spectral flattening beyond the tail break, which is very rare to report together in a GRG. This rare and remarkable system offers a unique laboratory for probing the interplay between active galactic nucleus activity, star formation, and environmental effects in cluster-surrounded GRGs.
\end{abstract}



\section{Introduction}
\label{sec:intro}
Active galactic nuclei (AGN), in radio-loud mode, squirt out bipolar jets ranging from parsec (pc) scale to Mpc scale in size. 
The radio lobe becomes fainter over time due to the energy loss of relativistic electrons, primarily through synchrotron radiation and inverse Compton (IC) scattering to the cosmic microwave background (CMB). The rate of these energy losses depends on several factors. Synchrotron losses are mainly governed by the strength of the magnetic field, which determines the radiative cooling timescale of the electrons, while the physical size of the radio galaxy influences the characteristic age and dynamical timescales of the emitting plasma. In contrast, IC losses become increasingly significant at higher redshifts because of the $(1+z)^{4}$ dependence of the CMB energy density. However, if they are situated within galaxy clusters, the pressure of the hot and dense intracluster medium (ICM) can lead to confinement of these lobes and prevent expansion losses \citep{En01}. Thus, the cluster environment can significantly prolong the lifetime of radio sources by changing their energetics, keeping them detectable by sensitive radio telescopes for a longer duration of time \citep{Sa22}.

In a large number of galaxy clusters and groups, diffuse radio emissions have been observed, proving the presence of prevalent cosmic magnetic fields and relativistic particles within the large-scale structure of the Universe. Their typical size ranges from 100 kpc to a few Mpc, and they exhibit a variety of morphologies. The emission exhibits a power--law energy distribution with a steep spectral index ($\alpha \sim$1.3), similar to old and dying or revived RGs. Many times, the origin of cluster radio sources, such as revived sources, i.e., gently re-energized tails \citep[GReET;][]{Ga17} or phoenices \citep{En01}, are attributed to the fossil electron population from RGs present in the cluster field. Furthermore, RGs are known to influence the thermal and dynamical state of ICM through AGN feedback, particularly in the form of mechanical (or kinetic) feedback. This involves the injection of energy via radio jets and lobes, which inflate cavities (or bubbles) in the X-ray emitting gas, drive shocks, and generate turbulence, thereby preventing runaway cooling in cluster cores \citep[e.g.,][]{Mo17}. Among many examples of strong interaction of RGs with the cluster medium, a classic example is NGC 4869 in the Coma Cluster \citep{La20}. 

Jets of RGs often display distortions and asymmetries, primarily due to their interactions with the surrounding environment, particularly in dense regions such as galaxy clusters. The ICM plays a significant role in shaping these jets, leading to a variety of morphological features observed in RGs. Differences in the density of the ICM, the presence of shock waves propagating through it, and the relative motion between the ICM and RG contribute to the bending, twisting, and complex structural formations of the radio-emitting plasma \citep{Be79, Jo17, Ku24}. Such environmental influences can cause the jets to lose their initial collimation, resulting in irregular or disrupted morphologies. Additionally, the dynamic nature of the host AGN itself can further introduce distortions. Variability in AGN activity, including episodic jet emissions or reorientation of the jet axis over time, can significantly alter the observed structure of radio jets. Changes in jet velocity, power, or direction, linked to accretion processes in the AGN, also contribute to these morphological complexities. Consequently, RGs in cluster environments frequently exhibit bent or disrupted jets, highlighting the intricate interplay between the internal processes of AGNs and the external cluster environment.

\begin{table*}
 \caption{Data analysis parameters of J1007+3540}
    \centering
    \begin{tabular}{cccccccc}
    \hline
        Telescope/Survey& Band  &Frequency (MHz)& Robust & Tapered ($k\lambda$)&RMS ($\mu$Jy beam$^{-1}$)& Resolution ($''$)& Position angle ($^{\circ}$) \\  \hline
        uGMRT &3& 400 & 0 & -- & ~21 & 9.20 $\times$ 6.23 & 57.85\\ 
        uGMRT & 3&400 & 0 & 10 & ~30 & 21.55 $\times$ 17.30 & 48.35\\
         LoTSS DR2 &--& 144 & --0.5 & --  & ~80 & 6 $\times$ 6 & --\\
         LoTSS DR2 &--& 144 & --0.5 & -- & 112 & 20 $\times$ 20 & --\\
        \hline
    \end{tabular}
    \label{tab:data}
\end{table*}
Interestingly, some RGs exhibit not only one active phase but rather multiple episodes of jet activity \citep[e.g.][]{Su96, La99, Sc00a, Sc00b}. For such radio sources, how long does the active phase last? This is one of the unanswered questions in AGN studies. The possibility of nuclear activity duty cycles is closely related to the length of the active phase. Such duty cycles can only be identified if there are ways to detect previous nuclear activity for an extended period so that it may be recognized when a new cycle begins. Extended radio sources with large-scale lobes can serve as valuable indicators of episodic activity in AGN, as these lobes can remain detectable for a couple of $\sim$100 Myr at low radio frequencies (e.g. frequency ranges observed by Low-Frequency Array (LOFAR)
and upgraded Giant Metrewave Radio Telescope (uGMRT)), even after the central energy supply from the AGN has ceased. This extended visibility reflects the long synchrotron lifetimes of low-energy electrons in weak magnetic fields of a few $\mu$G, comparable to or slightly above the CMB-equivalent magnetic field strength ($\sim$3.2 $\mu$G at $z=0$). If a subsequent phase of AGN activity begins before the older lobes have fully faded, the presence of newly formed jets within the older, relic lobes can reveal this restarted activity. Observationally, such systems are characterized by a young, compact radio source embedded within a more diffuse, aged radio structure. One well-known example of an RG that shows evidence of restarted AGN activity is 3C 219, which has been extensively studied as a candidate for such episodic jet formation \citep{Cl91, Cl92, Pe94, Wo24}.

One of the most intriguing characteristics of restarted RGs, particularly double-double RGs (DDRGs), is the presence of inner radio lobes formed during subsequent phases of jet activity. Numerical simulations of restarting jet systems, along with theoretical studies on the properties of jet-driven cocoons, suggest that the low-density environment inside the cocoon is generally insufficient to produce strong shocks necessary for the formation of hotspots and radio lobes \citep{Cl92, Pe07, Ha13}. However, \citet{Ka00} proposed that after a sufficiently long period (on the order of tens of millions of years), the density within the cocoon may increase enough to support the formation of new lobes and hotspots once the jet flow is re-established following a period of inactivity. The large physical sizes of DDRGs are attributed to these extended time scales of episodic activity. Additionally, the relatively lower density inside the cocoon compared to the surrounding intergalactic medium explains the lower radio luminosity of these inner lobes. \citet{Sc00b} estimated the advance speeds of the inner hotspots in the range between $0.19c$ and $0.57c$ for the orientation of the sources of $45\degr \!\le\! \theta \!\le\! 135\degr$. For one of the restarted sources, B1834$+$620 \citep{Sc00a, Sc00b}, in the sample of seven DDRGs, the advance speed of the inner hotspots is estimated as $0.298c \pm 0.006c$ using the orientation of the source, $\theta = 85.3\degr \pm 1.5\degr$.

In this paper, we report a giant-sized RG, \src{}, with an episodic feature in its radio morphology. The source is located near the centre of the galaxy cluster WHL J100706.4+354041. Throughout the paper, we have used the following cosmological parameters from the final full-mission \textit{Planck} measurements of the cosmic microwave background anisotropies: $H_0 = 67.4$ km s$^{-1}$ Mpc$^{-1}$, $\Omega_m = 0.315$ and $\Omega_{vac} = 0.685$ \citep{Ag20}. Throughout the paper, we adopted $S \propto \nu^{-\alpha}$ (where $S$ is the flux density and $\alpha$ is the spectral index) convention for spectral index measurements.

\section{Data and analysis}
\label{sec:observation}
This section discusses the LOw-Frequency ARray \citep[LOFAR;][]{va13} data from where the reported source is identified and the GMRT \citep[GMRT;][]{Sw91} observation and data reduction analysis procedures.

\subsection{LoTSS DR2 survey data}
\label{subsec:lofar}
We used the LOFAR Two-metre Sky Survey second data release (LoTSS DR2) \citep{Sh22} at 144 MHz to search for giant radio galaxies (GRGs). LoTSS DR2 covers 4396228 radio sources, including the LoTSS DR1 \citep{Sh19} source catalogue. We filtered the sources with an angular size $\geq 4'$. This source selection method is also discussed in detail in \citet{Pa23}. From the filtered catalogue, we conducted a detailed manual inspection to identify RGs that are associated with known galaxy clusters, specifically focusing on the WHL cluster catalogue \citep{We12, We15}, the the Dark Energy Spectroscopic Instrument’s
Legacy Imaging Surveys \citep[DESI LS DR9;][]{Sc21} cluster catalogue \citep{We24}, and the Abell cluster catalogue \citep{Ab89}. Through this targeted search, we identified over 20 RGs exhibiting irregular and distorted radio morphologies in proximity to the cluster centers, suggesting possible environmental influences such as interactions with the ICM.
For a pilot study, we selected one particularly interesting source, a GRG J1007+3540. This source is located near the central region of a WHL cluster and displays the most compelling evidence of strong jet--ICM interactions among the candidates. Its large angular size, distorted radio morphology, and asymmetries in the jet and lobe structures make it an ideal case study for investigating the impact of the dense cluster environment on the evolution and dynamics of radio jets.

We checked the detectability of the source presented in the current paper in all available large-area radio sky surveys. We visually checked surveys like the Very Large Array (VLA) Faint Images of the Radio Sky at Twenty Centimeters (FIRST) survey at 1400 MHz \citep{Be95, Wh97}, the National Radio Astronomy Observatory (NRAO) VLA Sky Survey (NVSS) at 1400 MHz \citep{Co98}, the TIFR GMRT Sky Survey Alternative Data Release 1 (TGSS ADR1) at 150 MHz \citep{In17}, GaLactic and Extragalactic All-sky Murchison Widefield Array (GLEAM) at 72--231 MHz \citep{Hu17}, the Westerbork Northern Sky Survey (WENSS) at 326 MHz \citep{Re97}, and the VLA Sky Survey (VLASS) at 3 GHz \citep{La2020}. The source J1007+3540 is detected in TGSS and NVSS, but no extended and diffuse emission has been found. Due to the coarser resolution (45\arcsec in the NVSS and 25\arcsec in the TGSS) in comparison to the LoTSS DR2 at 144 MHz and the lower sensitivity (0.45 mJy beam$^{-1}$ in the NVSS and 3.5 mJy beam$^{-1}$ in the TGSS), the NVSS and TGSS surveys could not detect any extended and diffuse emission from the source J1007+3540. However, because of having much better sensitivity (83 $\mu$Jy beam$^{-1}$) and resolution (6\arcsec), LoTSS DR2 at 144 MHz has detected extended and diffuse emissions in J1007+3540. In the LoTSS DR2 catalogue \citep{Sh22}, this object has the standard LOFAR identifier ILTJ100707.29+354049.9. Throughout this article, we refer to the object as J1007+3540 for convenience.

\subsection{uGMRT data and analysis}
\label{subsec:gmrt}
We used the uGMRT band 3 (250--500 MHz) receivers to carry out 4 hours of dedicated observation (Proposal code 45\_002; P.I.: S. Kumari) for the source (J1007+3540) presented in the current article. The data was recorded with 2048 frequency channels and 10.2 s integration time in 4 scans. 3C 147 and 3C 286 were used as flux calibrators and observed for 5 min and 4 min at the beginning and end of the observation, respectively. 1021+219 was used as the phase calibrator and was observed four times throughout the run for five minutes per scan. The uGMRT RFI filter \citep{Bu22, Bu23} was set to be ON during the observation. We thoroughly examined the data to identify and remove bad time stamps, bad antennas, and segments affected by radio frequency interference (RFI). Following these quality checks, initial calibration steps, including flux density and bandpass calibration, were performed. All bad data were subsequently flagged and removed using standard data reduction routines. For initial flagging, we applied the automated flagging task \textsc{tfcrop} and also used the CASA pipeline-cum-Toolkit for upgraded Giant Metrewave Radio Telescope data reduction (CAPTURE) \citep{Ka21}. The absolute flux density calibration was carried out using the flux scale defined by \citet{Pe17}. The \textsc{casa} task \textsc{tclean} is used to perform wide-band 3D imaging. It employs \texttt{gridder = `widefield'}, Briggs weighting, with a robustness parameter set to 0 (\texttt{robust = 0}), and two Taylor series expansion terms (\texttt{nterms = 2}, representing \texttt{tt0} and \texttt{tt1}). Finally, tasks \textsc{gaincal} and \textsc{applycal} in \textsc{casa} were used for amplitude and phase self-calibration. The data analysis parameters are tabulated in Table \ref{tab:data}. 
\begin{figure*}
\vbox{
\centerline{
\includegraphics[width=18cm,origin=c]{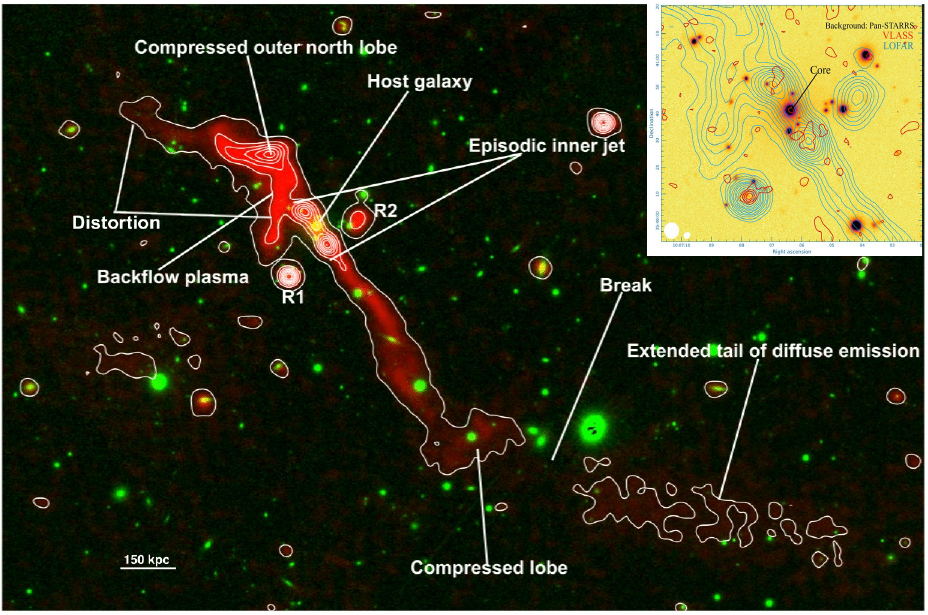}
}
}
\caption{LoTSS DR2 image of \src{} at 144 MHz (in contour) superimposed with Pan-STARRS1 optical r-band image \citep{Ch16}. The contour levels are at 0.24, 1.3, 2.6, 4.3, 5.2, 6.0, 7.0, 8.0, 8.5, 9.0, 12.2, 20.6, 34.8 mJy beam$^{-1}$. The lowest contour level of \src{} is set at 3$\sigma$, where $\sigma = 80$ $\mu$Jy beam$^{-1}$ represents the mean RMS noise measured around the source. The red colour represents the LoTSS DR2 image at 144 MHz, and the green colour represents the Pan-STARRS1 optical image. The upper right-side inset represents the overlaid image of J1007+3540 with LoTSS DR2 at 144 MHz (green contour), VLASS at 3 GHz (red contour), and optical in the background.}
\label{fig:LOFAR_high}
\end{figure*}

\begin{figure}
\vbox{
\centerline{
\includegraphics[width=9.2cm,origin=c]{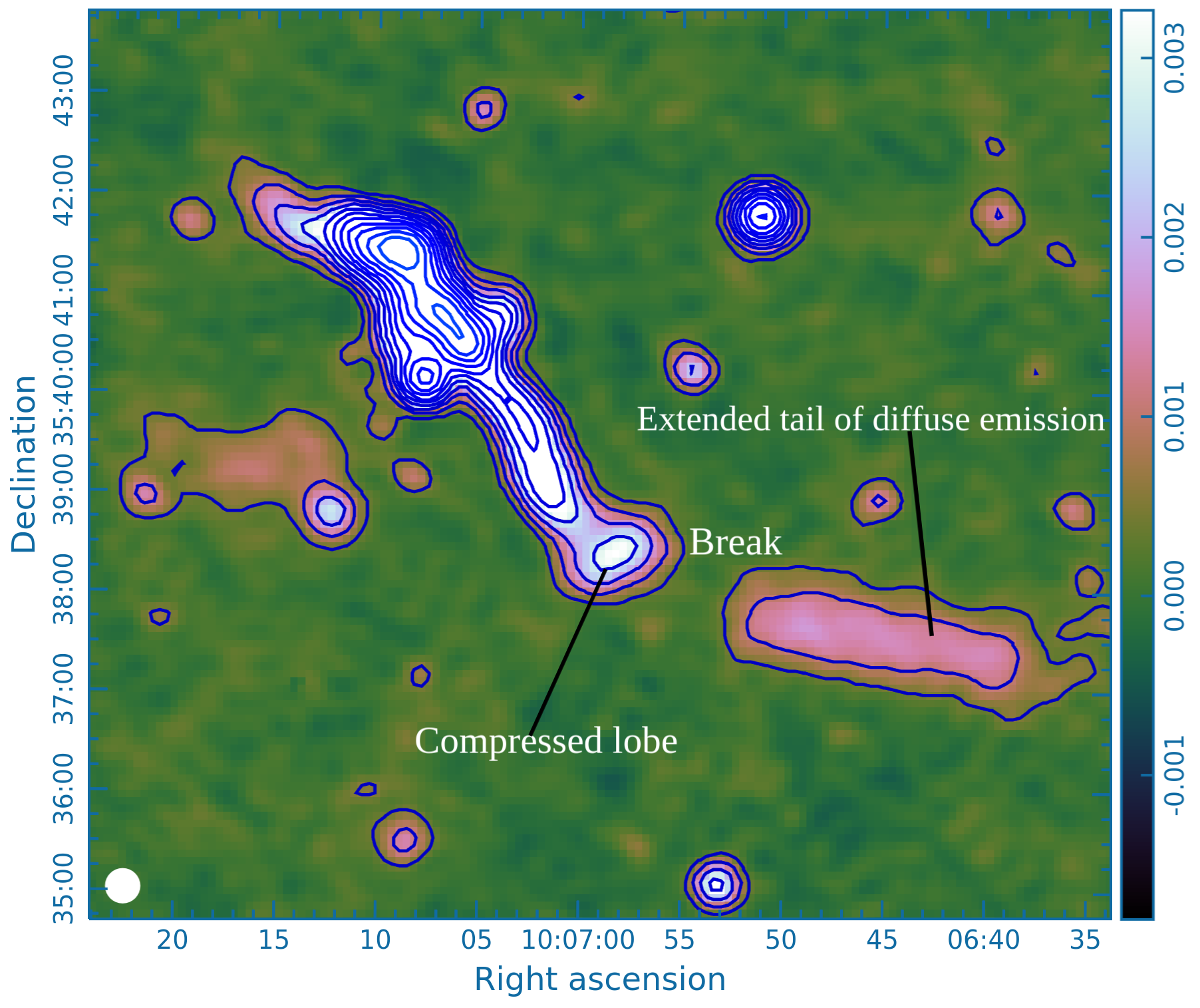}
}
}
\caption{LoTSS DR2 image at 144 MHz at 20$''$ resolution (low-resolution beam). The contour levels are at 0.42, 0.87, 2.07, 2.85, 3.79, 4.91, 6.26, 7.88, 9.83, 12.2, 15.0, 18.3, 22.4, 28.6, 36.3 mJy beam$^{-1}$. The lowest contour level is set at 3$\sigma$, where $\sigma = 0.14$ mJy beam$^{-1}$ represents the mean RMS noise measured around the source. The synthesized beam of \src{}, shown as a white circle in the bottom-left corner of the image, has an angular resolution of $20\arcsec \times 20\arcsec$. The colour scale represents the specific intensity of the source in Jy~beam$^{-1}$.}
\label{fig:LOFAR_low}
\end{figure}

\begin{figure*}
\vbox{
\centerline{
\includegraphics[width=15cm,origin=c]{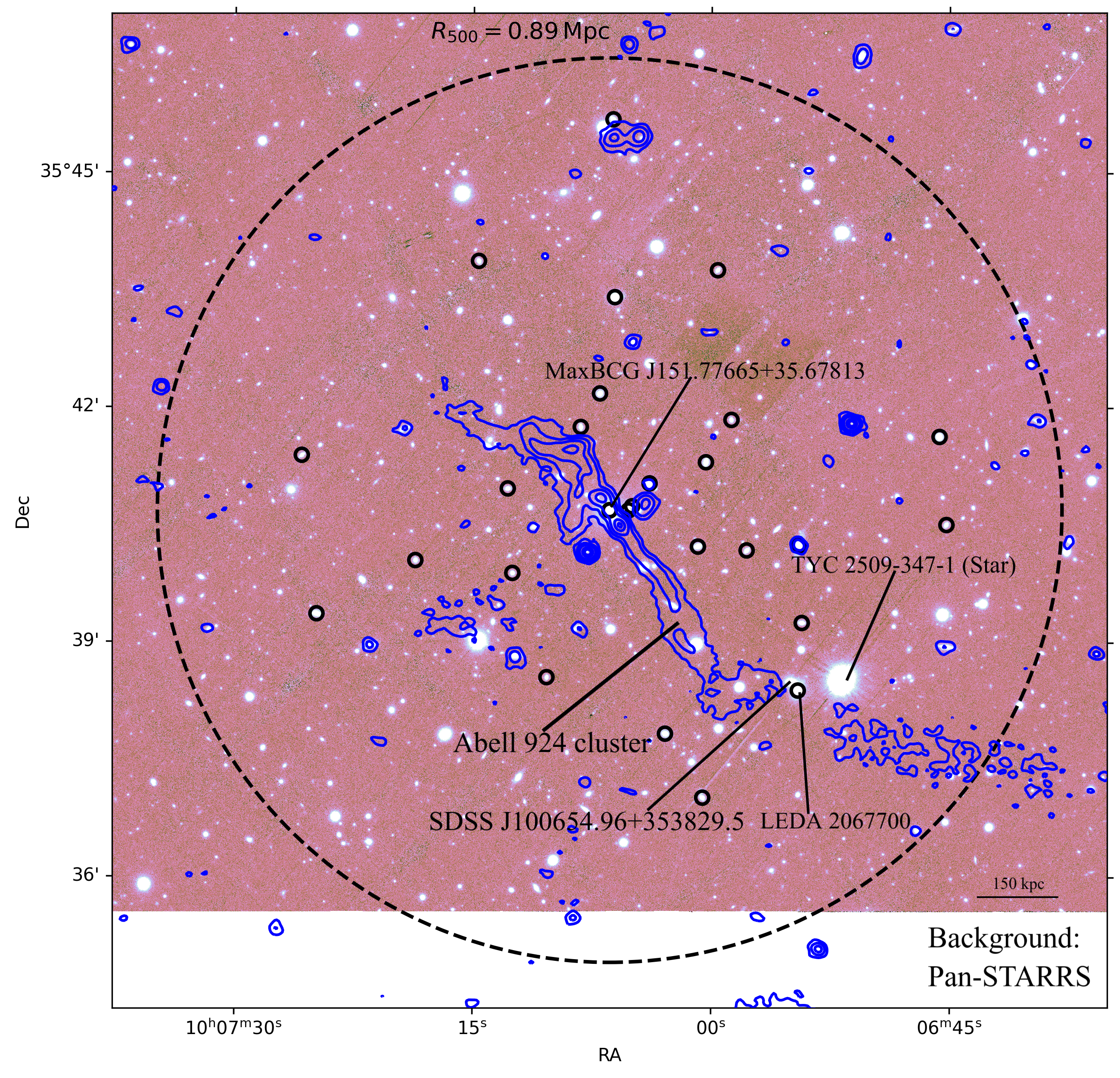}
}
}
\caption{Plot of galaxies resides within the cluster radius of R$_{500}$ = 0.89 Mpc. The large black dashed circle is the locus of all points whose sky-projected distance from the cluster centre is R$_{500}$. Small black circles overplotted on the LoTSS DR2 image (in blue contour) represent optical galaxies within the R$_{500}$. The contour levels are at 0.23, 0.63, 1.60, 2.83, 4.40, 5.34, 6.40, 7.60, 8.96, 10.50, 13.0 mJy beam$^{-1}$. The background image is the optical image taken from the Pan-STARRS \citep{Ch16}.}
\label{fig:cluster_galaxies_plot}
\end{figure*}
 \subsection{Optical SED analysis}
\label{subsec:SED1}
To investigate the physical properties of the host galaxy and its central AGN, we performed a multi-component spectral energy distribution (SED) fitting of J1007+3540 using the \texttt{AGNfitter} code \citep{Ca16, Ma24}. The fitting was based on broadband photometric data covering the mid-infrared (MIR), near-infrared (NIR), and optical domains. We used the Wide-field Infrared Survey Explorer \cite[WISE;][]{Wr10, Cu21} and Two Micron All Sky Survey \citep[2MASS;][]{Hu12} catalog to study the infrared properties of the host galaxy. The WISE uses the W1, W2, and W3 bands at wavelengths of 3.4 $\mu$m, 4.6 $\mu$m, 12 $\mu$m, and 22 $\mu$m, respectively. The 2MASS uses J, H, and Ks bands at 1.235 $\mu$m, 1.662 $\mu$m and 2.159 $\mu$m, respectively. For the study of the optical domain, we utilised the Panoramic Survey Telescope and Rapid Response System \citep[Pan-STARRS1;][]{Ch16} photometric data, which spans the $g$, $r$, $i$, $z$ and $y$ bands in the wavelength range of 0.4 to 1.0 $\mu$m. The IR and optical photometric data collected from these surveys are tabulated in Table \ref{tab:survey_band_flux_error}. \texttt{AGNfitter} models the observed SED by decomposing it into four physical components: (1) stellar emission from the evolved stellar population of the host galaxy \citep{Br03}, (2) a starburst component representing the thermal emission from dust heated by young stars \citep{Sc18}, (3) the AGN accretion disk or ``big blue bump'' \citep[BBB;][]{Te21}, and (4) the emission from the dusty torus surrounding the AGN \citep[e.g.][]{St16}, which reprocesses UV/optical radiation into the infrared. The fitting was carried out using a Markov Chain Monte Carlo (MCMC) approach to sample the posterior distributions of the model parameters, allowing for robust statistical estimation of physical quantities and their associated uncertainties.
\section{Multi-wave band characteristics of the source J1007+3540}
\label{sec:multiband}
\subsection{Optical counterparts and the cluster environment}
\label{subsec:optical1}
The source J1007+3540 is found to be hosted by the brightest cluster galaxy (BCG), MaxBCG J151.77665+35.67813 (also named as PGC~2068293), with the r-band magnitude of r$_{mag} = 16.13$ \citep{Ha10, We12, Ahu20}. 
MaxBCG J151.77665+35.67813 is at a spectroscopic redshift of $z = 0.14335 \pm 0.00002$ \citep{Cu12}. This BCG is located near the optical centre of the WHL~J100706.4+354041 cluster ($z = 0.14392 \pm 0.00046$; \citealt{Bl17, Ki21}), where the cluster centre is defined by the optical centroid derived from the positions of spectroscopically confirmed member galaxies. \citet{We15} reported the radius of the WHL J100706.4+354041 cluster as R$_{500}=0.89$ Mpc. Here, R$_{500}$ is defined as the largest radius within which the overdensity ($\Delta_{cr}$) exceeds 500, where $\Delta_{cr}= \frac{\rho}{\rho_{cr}}$, $\rho_{cr} = \frac{3H(z)^2}{8 \pi G}$ represents the critical density of the Universe at redshift $z$, determined by the Hubble parameter at redshift ($z$), $H(z)$. \citet{We15} also reported the richness factor of the cluster to be (R$_{Cl}$) as 40.6. The total mass enclosed within a radius of R$_{500}$, denoted as $M_{500}$, is noted as $2\times10^{14} M_{\odot}$ using the information of cluster richness R$_{Cl}$ in richness--mass scaling correlation \citep[see equ.\ref{eq:mass-cluster};][]{An10}. The total mass ($M_{500}$) includes contributions from dark matter, the ICM, and galaxies, with the dark matter component dominating the total mass budget.
\begin{equation}
\label{eq:mass-cluster}
    \log M_{500}=(1.08\pm0.02)\log \textrm{R}_{Cl}-(1.37\pm0.02)
\end{equation}

\begin{table*}
\centering
\caption{Optical and infrared photometric data used in the SED fitting}
\label{tab:survey_band_flux_error}

\begin{tabular}{cccccc}
\hline
Survey & Band &Wavelength ($\mu$m)& Flux Density (mJy) & Flux Density Uncertainty (mJy) \\
\hline
Pan-STARRS1 (PS1) \citep{Ch16}& $g$ & $0.4832$&$0.365$ & $0.013$ \\
                    & $r$ & $ 0.6188$ &$0.808$& $0.010$ \\
                    & $i$ & $0.7525$ &$1.136$& $0.009$ \\
                    & $z$ & $0.8669$ &$1.449$& $0.010$ \\
                    & $y$ & $0.9622$ &$2.139$& $0.025$ \\\hline
2MASS \citep{Hu12}   & $J$   & $1.2376$ &$2.14$& $0.1$ \\
          & $H$   & $1.6476$ & $2.78$&$0.16$ \\
          & $K_s$ & $2.1621$ & 2.76&$0.19$ \\\hline
WISE \citep{Wr10, Cu21}     & $W1$  & 3.4&$2.21$ & $0.02$ \\
          & $W2$  & 4.6&$1.47$ & $0.02$ \\
          & $W3$   & 12&$27.2$ & $0.22$ \\
\hline
\end{tabular}
\end{table*}

\subsection{Results from LOFAR at 144 MHz}
\label{subsec:J1007+3540}

The high-resolution LoTSS DR2 image at 6$''$ of J1007+3540 is shown in Fig. \ref{fig:LOFAR_high}. In this figure, the LoTSS DR2 radio image is overlaid with the Pan-STARRS1 optical image. The upper right-side inset in Fig. \ref{fig:LOFAR_high}, represents the overlaid image of J1007+3540 with LoTSS DR2 at 144 MHz (green contour), the VLASS image at 3 GHz (red contour), and optical from Pan-STARRS1 in the background. The radio core is detected in VLASS image (marked by `Core' in the inset of Fig. \ref{fig:LOFAR_high}) at 2$\sigma$ lowest contour level ($\sigma=75\mu$Jy beam$^{-1}$), coincides with the optical counterpart (marked by the `host galaxy' in Fig. \ref{fig:LOFAR_high}). The RG, J1007+3540, is found to be a GRG with a projected linear extension ($D_p$) of 1.45 Mpc (measured using $D_p=\frac{\theta\times D_{co}}{(1+z)}$, where $D_{co}$ is the comoving distance to J1007+3540, $\theta$ is the largest angular size (LAS = 552\arcsec) of J1007+3540, and $z = 0.14335 \pm 0.00002$ \citep{Cu12} is the redshift). The source possesses a distorted northern outer lobe and distorted backflow in the structure along the southeastern side, and an extended tail of diffused emission with a compressed lobe and the break elongated towards the southwest side of the structure of the galaxy. We have looked for the optical counterparts of the radio-bright sources labelled R1 and R2, located near the inner south lobe (see Fig. \ref{fig:LOFAR_high}). The point-like source R1 coincides with a weak optical counterpart located at 10:07:07.84 +35:40:09.3, visible in Pan-STARRS1, DESI \citep{Sc21}, and the Hyper Suprime-Cam Legacy Archive \citep[HSCLA;][]{Ta21}. The photometric redshift for R1 is $z = 0.588\pm0.411$ \citep{Sc21}, which is distinct from the spectroscopic redshift of J1007+3540 ($z = 0.14335 \pm 0.00002$). As no spectroscopic redshift information is available for R1, and the photometric redshift of R1 is prone to large error, R1 is likely a background point source unrelated to J1007+3540. The source R2 is relatively fainter and slightly extended in the radio image compared to R1. The centre of R2 coincides with a faint optical counterpart at 10:07:04.10 +35:40:44.2, visible in HSCLA, but no redshift information is available for it. The low-resolution LoTSS DR2 map with a 20\arcsec $~$beam (see Fig. \ref{fig:LOFAR_low}) effectively captures the extended tail of diffuse emission, revealing a compressed lobe and a discontinuity (a break) in the extended tail of diffuse emission. J1007+3540 is not catalogued in the GRGs list of \citet{Oe23}. The reason for this may be because if we measure the size of J1007+3540 before the break, excluding the faint, extended tail of diffuse emission beyond the apparent break, the projected linear extension of \src{} is less than the 700 kpc threshold that is typically used to define a GRG \citep{Ku18, Da20a, Da20b, Bh24}. \citet{Oe23} focused on identifying GRGs with well-defined, continuous radio structures through a combination of automated selection and visual inspection. In cases where diffuse, low surface brightness extensions were fragmented or poorly connected to the main lobes, these structures were possibly not included in the overall size measurement in \citet{Oe23} to maintain a robust classification. Therefore, it is possible that the extended and diffuse tail of J1007+3540 was not included in the analysis of \citet{Oe23}, leading to its omission from their GRG catalogue despite the true physical size of the source exceeding the GRG threshold.

We have overplotted the distribution of optical galaxies within the cluster radius (R$_{500} = 0.89$ Mpc) on the LoTSS DR2 high-resolution image (6$''\times 6''$) at 144 MHz (in blue contour; see Fig. \ref{fig:cluster_galaxies_plot}). Each galaxy position is indicated by a black circle. \citet{We15} identified thirty-two optical galaxies (see Table \ref{Tab:WHL} for WHL cluster properties) projected within the WHL J100706.4+354041 cluster radius of R$_{500} = 0.89$ Mpc, using a redshift offset of $\delta z = 0.05$. In our analysis (Fig. \ref{fig:cluster_galaxies_plot}), we adopt a narrower redshift offset of $\delta z = 0.01$, resulting in a total of twenty-eight identified galaxies. The distribution of these galaxies appears concentrated toward the cluster centre.

As can be seen in Fig. \ref{fig:cluster_galaxies_plot}, TYC 2509-347-1 is a star located close to the break of the southern extended tail of diffuse emission. With a spectroscopic redshift of $z=0.05639$, SDSS J100654.96+353829.5, which is also located after the compressed lobe prior to the break of the extended tail of diffuse emission, exhibits a significantly lower redshift, indicating that it is likely a foreground galaxy and not physically associated with the WHL cluster.
At an angular distance of 1.63 arcmin from the centre of J1007+3540, there is an Abell 924 cluster at a photometric redshift of $z=0.0908$, where the host galaxy and the sources within the R$_{500}$ radius of the WHL J100706.4+354041 cluster have redshifts in the range of $z=0.13951$ to $z=0.15125$. The significantly lower redshift of the Abell 924 cluster compared to J1007+3540 suggests that these two are likely not related and are close only in an angular sense.

\begin{table}
\begin{center}

\caption{Properties of WHL J100706.4+354041 cluster obtained from \citet{We15}.}
\begin{tabular}{lllll}

\hline
 $\rm R_{500}$ & $\rm R_{L*}$ & $\rm N_{500}$ & $\rm M_{500}$ & V  \\  
  (Mpc) & &  & ($\rm M_{\odot}$) & ($\rm Mpc^{3}$) \\  
  (1) & (2) & (3) & (4) & (5)   \\ \hline 
0.89 & 40.6 & 32 & 2 $\times$ $10^{14}$ & 4.4 \\
\hline
    \\
\end{tabular}
\\
NOTE. (1) $\rm R_{500}$: radius of the galaxy cluster; (2) $\rm R_{L*}$: richness of the galaxy cluster,   defined as $\rm R_{L*} = \frac{\rm L_{500}}{L_*}$, where L$_*$ is the characteristic luminosity of galaxies in the r band and $\rm L_{500}$ the total r-band luminosity within the radius of $\rm R_{500}$; (3)$\rm N_{500}$: number of galaxies observed within $\rm R_{500}$; (4) $\rm M_{500}$: mass within $\rm R_{500}$; (5) V: volume of cluster within $\rm R_{500}$. 
\label{Tab:WHL}
\end{center}
\end{table}

\begin{figure*}
\vbox{
\centerline{
\includegraphics[width=9.55cm,origin=c]{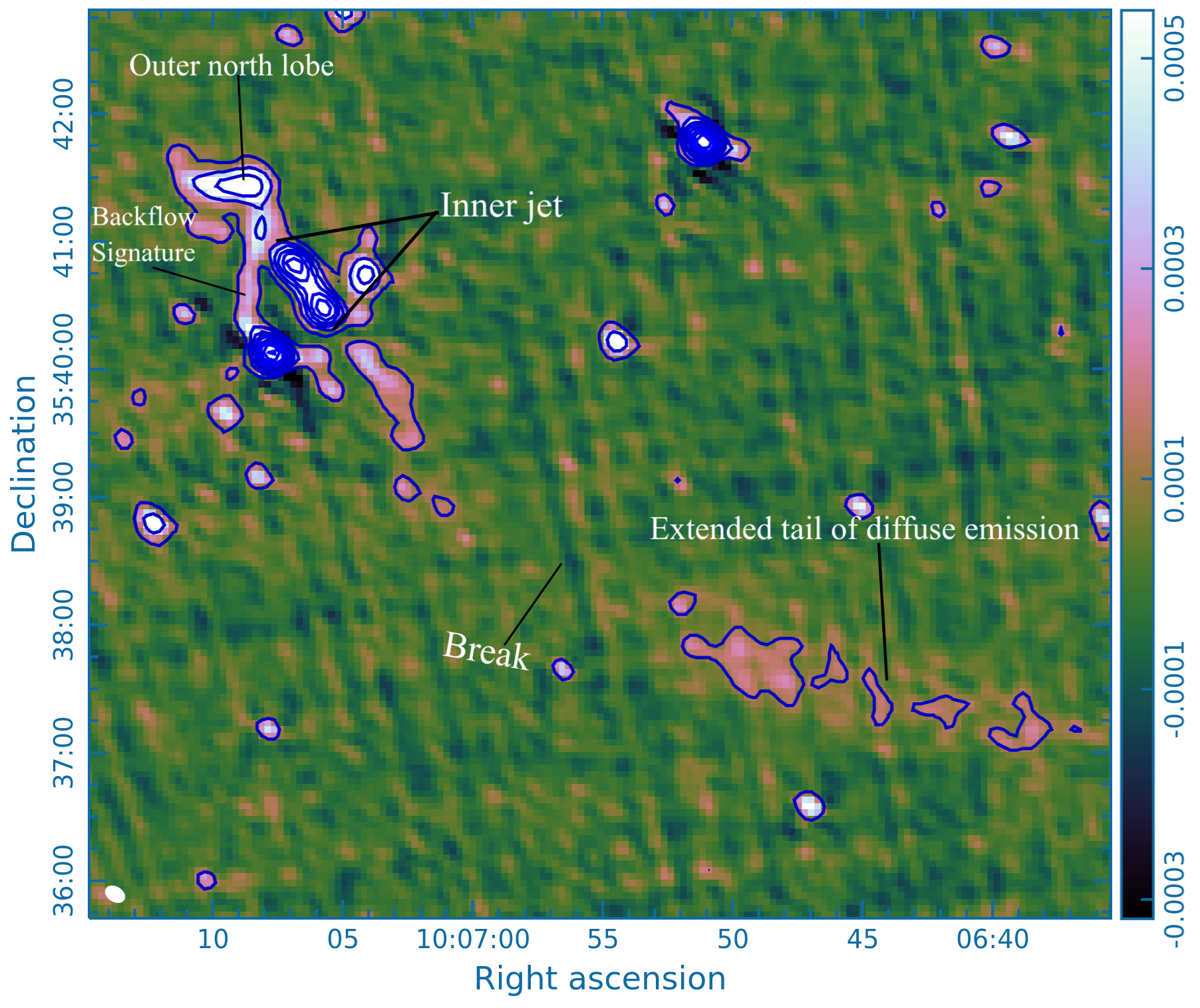}
\includegraphics[width=9.3cm,origin=c]{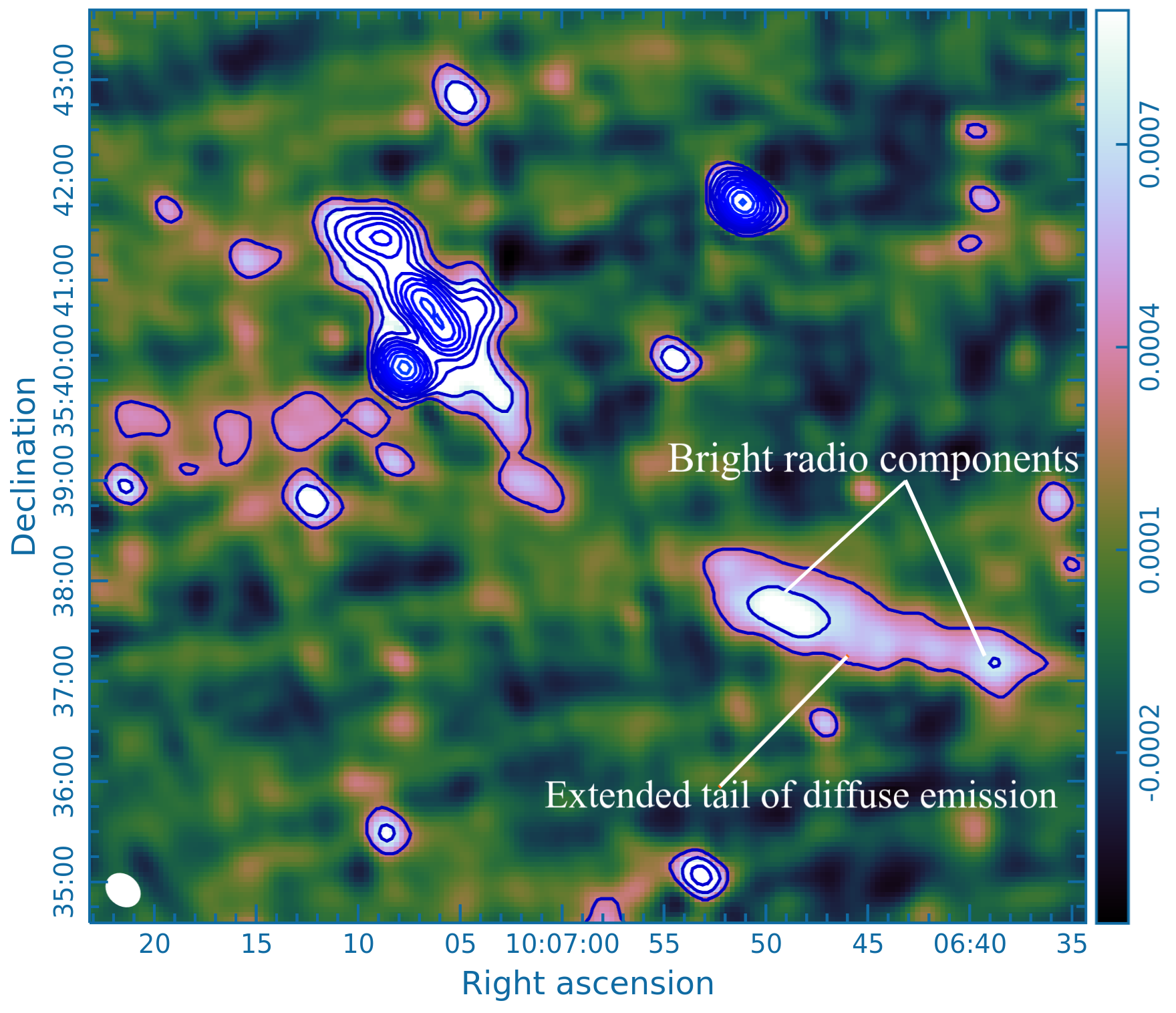}
}
}
\caption{Left: High-resolution uGMRT image of J1007+3540 at 400 MHz (with resolution of $9.20''\times 6.23''$; position angle = 57.85$^{\circ}$). The contour levels are at 0.06, 0.15, 0.28, 0.44, 0.66, 0.96, 1.36, 1.91, 2.64, 3.64, 4.98, 6.79 mJy beam$^{-1}$. The lowest contour level is at 3$\sigma$, where $\sigma=21 \mu$Jy beam$^{-1}$ is the mean RMS around the source. Right: uGMRT image of J1007+3540 at 400 MHz by applying uv taper (with synthesized beam of resolution $21.55'' \times 17.3''$; position angle: 48.35$^{\circ}$). The contour levels are at 0.09, 0.19, 0.32, 0.47, 0.67, 0.91, 1.22, 1.60, 2.07, 2.67, 3.42, 4.35, 5.51, 6.97, 8.78 mJy beam$^{-1}$. The lowest contour level is at 3$\sigma$, where $\sigma=30 \mu$Jy beam$^{-1}$ is the mean RMS around the source. The colour scale represents the intensity of the source in Jy~beam$^{-1}$.}
\label{fig:GMRT_low+high}
\end{figure*}

\subsection{Result of uGMRT observation}
\label{subsec:result}
We present the high-resolution uGMRT band~3 map of J1007+3540 in the left panel of Fig. \ref{fig:GMRT_low+high}. The synthesized beam has been indicated by a white ellipse in the lower-left corner of the image. The beam size is $9.20''\times 6.23''$; position angle = 57.85$^{\circ}$ at a frequency of 400 MHz (imaged for the full band). We have achieved an average RMS of 21 $\mu$Jy beam$^{-1}$. 
 To enhance the detection of diffuse emission from J1007+3540, we applied a uv taper of 10 k$\lambda$ and a weight parameter, robust = 0. This produced a low-resolution map with a beam of 21.55$''$ $\times$ 17.30$''$; position angle: 48.35$^{\circ}$ (see the right panel of Fig. \ref{fig:GMRT_low+high}) and we achieved an RMS near the centre of the uGMRT pointing of target as 30 $\mu$Jy beam$^{-1}$. As can be seen in the high-resolution and uv-tapered uGMRT images (see Fig. \ref{fig:GMRT_low+high}), the uGMRT was unable to map the compressed lobe just before the break in J1007+3540 compared to the emission captured in the high as well as in low-resolution LoTSS DR2 images at 144 MHz (see the compressed lobe in Figs. \ref{fig:LOFAR_high} and \ref{fig:LOFAR_low}).

\subsection{Radio multi-band properties of J1007+3540}
\label{sub:radio_pro}
\subsubsection{Flux density measurements}
\label{sub:flux}
The flux densities of different parts of the structure (e.g., inner north lobe, inner south lobe, outer north lobe, backflow structure, and extended tail of diffuse emission) of J1007+3540 at 144 MHz and 400 MHz are tabulated in the 1st and 2nd columns of Table \ref{tab:radio_high}.
Furthermore, the corresponding spectral indices between these two frequencies are calculated and tabulated in the 3rd column of Table \ref{tab:radio_high} for each part of J1007+3540. The uncertainty in flux density measurements for LoTSS DR2 at 144 MHz is estimated to be $\sim$10\% \citep{Sh22}. We computed the uncertainty ($\sigma_S$) associated with the measured 400 MHz uGMRT flux density using the relation $\sigma_S = \sqrt{\sigma^2 \cdot N_{\mathrm{beam}} + (f_{\mathrm{uGMRT}} \cdot S)^2}$, where $\sigma$ is the mean RMS in the field of J1007+3540, $S$ is the total integrated flux density of \src{}, and $N_\mathrm{beam}$ is the number of beams within 3$\sigma$ contours of the image. We assumed $\sim$10\% absolute flux scale uncertainty for the uGMRT ($f_\mathrm{uGMRT}$) band 3 image.

\begin{table*}
 \caption{Radio properties of different morphological parts of J1007+3540 }
    \centering
    \begin{tabular}{lccccccc}
\hline
Parts of the source & $S_{144}$ & $S_{400}$ & $\alpha$ & $L_{144}$ & $B_\mathrm{eq}$ & $\nu_\mathrm{br}$ & $t_\mathrm{rad}$  \\
 & (mJy) & (mJy) & ($\pm$0.1) & ($\times10^{24}$ W Hz$^{-1}$) & ($\mu$G) & (MHz) & (Myr)  \\
\hline
Inner north lobe & 17.6 $\pm$ 1.76 & 6.32 $\pm$ 0.32 & 1.0 & 1.03 & 2.16 & 513 & 139  \\
Inner south lobe & 17.7 $\pm$ 1.77 & 6.19 $\pm$ 0.31 & 1.0 & 1.04 & 2.34 & 493 & 142  \\
Outer north lobe & 91.1 $\pm$ 9.11 & 10.14 $\pm$ 0.51 & 2.1 & 6.21 & 1.65 & 155 & 242  \\
Feedback/backflow structure & 25.5 $\pm$ 2.55 & 2.30 $\pm$ 0.12 & 2.3 & 1.80 & 1.82 & 138 & 262  \\
Extended south tail & 26.8 $\pm$ 2.70 & 12.3 $\pm$ 0.60 & 0.8 & 1.54 & 0.64 & 457 & 100  \\
\hline
\end{tabular}
    \label{tab:radio_high}
\end{table*}

\subsubsection{Spectral index measurements}
\label{sub:spec_index}
The high-resolution data at 144 and 400 MHz, where the inner jet and backflow features are resolved, allow for precise spectral index measurements (see the 3rd column in Table \ref{tab:radio_high}). The error in the spectral index measurements between LoTSS DR2 (144 MHz) and uGMRT (400 MHz) images is

\begin{equation}
	\Delta\alpha=\frac{1}{\log\frac{\nu_{1}}{\nu_{2}}\times\ln{10}}\sqrt{\left(\frac{\Delta S_{1}}{S_{1}}\right)^{2}+\left(\frac{\Delta S_{2}}{S_{2}}\right)^{2}},
	\label{equ:equ2}
\end{equation}
where $\nu_{1, 2}$ and $S_{1, 2}$ represent the LoTSS DR2 (144 MHz) and uGMRT (400 MHz) frequencies and flux densities, respectively. The uncertainty in the measured spectral index between LoTSS DR2 and uGMRT (400 MHz) images using equation \ref{equ:equ2} is $\Delta\alpha = 0.1$. The inner lobes show spectral indices of 1.0 for the northern and southern lobes, respectively. The outer north distorted lobe exhibits a steeper spectral index of 2.1, while the distorted backflow alone has an even steeper spectral index of 2.3. In contrast, the large extended tail of diffuse emission is not well detected in the high-resolution images at these two frequencies (see Fig. \ref{fig:LOFAR_high} and the left panel of Fig. \ref{fig:GMRT_low+high}) due to the loss of diffuse emission. Therefore, its spectral index was calculated using the low-resolution (144 MHz LoTSS DR2 image at $20 \arcsec \times 20 \arcsec$ and 400 MHz tapered image of uGMRT with  $21.55''\times 17.3''$) images, where the emission is better captured (see Fig. \ref{fig:LOFAR_low} and the right panel of Fig. \ref{fig:GMRT_low+high}). The spectral index for the large extended tail after the break is 0.8. As discussed in Section \ref{subsec:result}, the uGMRT is unable to map the compressed lobe just before the break. We measured an upper limit for the spectral index of the compressed lobe using the LoTSS DR2 at 144 MHz in low-resolution (with flux of 3.7 mJy) and $3\sigma$ flux of uv-tapered uGMRT at 400 MHz (0.09 mJy). This gives the spectral index of $\alpha \lesssim 3.6$. This indicates that the spectrum of this region is possibly ultra-steep and populated with old fossil electrons. Despite the high sensitivity of the uGMRT Band 3 map (RMS$ =21 ~\mu$Jy beam$^{-1}$), this feature was not detected in either the full-band high-resolution or the uv-tapered low-resolution images, even with about 3 hours of on-source integration. The spatial distribution of spectral indices for the various parts of J1007+3540 is illustrated in the spectral index maps (discussed in Section \ref{sub:spec_index_map}). The high-resolution spectral index map (see Fig. \ref{fig:spec_map_high}) provides details of the spectral index distribution of the inner jet, outer northern lobe, and backflow features, while the low-resolution spectral index map (see Fig. \ref{fig:spec_age_low}) highlights the spectral index distribution for the large extended tail of diffuse emission. 
\subsubsection{Spectral index map}
\label{sub:spec_index_map}
\begin{figure}
\vbox{
\centerline{
\includegraphics[width=9.0cm,origin=c]{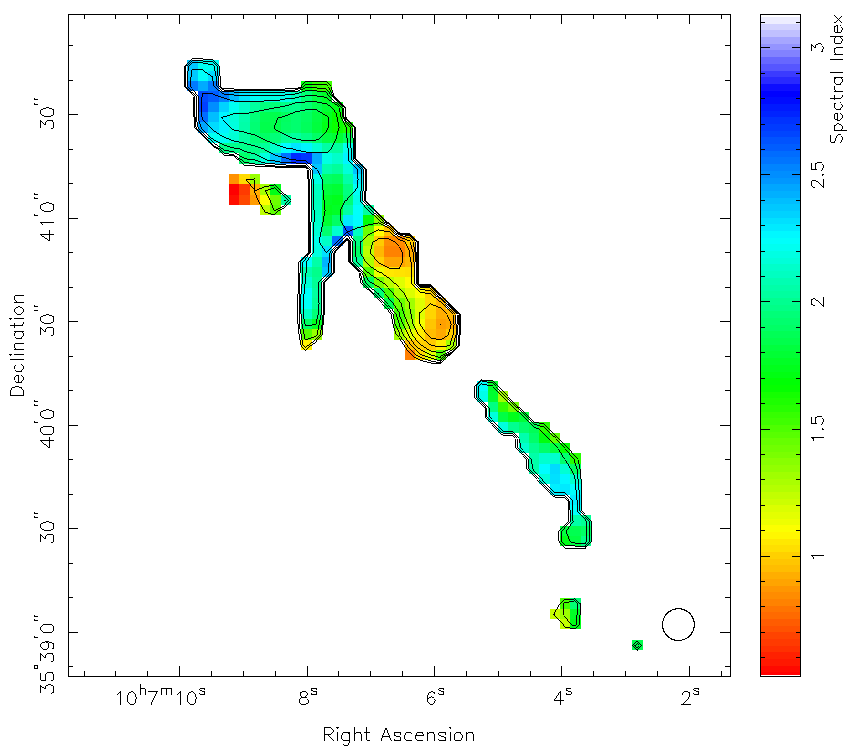}
}
}
\caption{Spectral index map of J1007+3540 using LoTSS DR2 at 144 MHz and uGMRT at 400 MHz with $9.20\arcsec \times 9.20\arcsec$ resolution. The contour plot presents the LoTSS DR2 map of \src{} at 144 MHz. The contour levels are at 3$\sigma \times (\sqrt{2})^{n}$, where $n$ = 0, 1, 2, 3,....}
\label{fig:spec_map_high}
\end{figure}

We created the high-resolution spectral index map (see Fig. \ref{fig:spec_map_high}) of \src{} between 144 and 400 MHz. In this map, we used the LoTSS DR2 image at 6\arcsec and the full resolution uGMRT image ($9.20\arcsec \times 6.23\arcsec$). We convolved the LoTSS DR2 and uGMRT images to a common resolution of $9.20\arcsec \times 9.20\arcsec$. In this map, the spectral indices of the inner lobes and outer northern lobe (with distorted backflow emission) vary from 0.85 to 1.1 and 1.9 to 2.5, respectively. Fig. \ref{fig:spec_age_low} represents the spectral index map of the extended tail of diffuse emission after the break in \src{} between 144 and 400 MHz. For this spectral index map, we utilized the LoTSS DR2 image at 144 MHz with an angular resolution of $20\arcsec \times 20\arcsec$. The uGMRT image in the full resolution ($9.20\arcsec \times 6.23\arcsec$) at 400 MHz is unable to map the southern extended tail of diffuse emission as observed in the LoTSS DR2 image at 144 MHz. Therefore, to detect the extended tail of diffuse emission, we applied a uv taper of 10 k$\lambda$ (equivalent to  22\arcsec) on uGMRT band 3 data during deconvolution. The uGMRT uv-tapered image captures the tail of extended diffuse emission (see the right panel image of Fig. \ref{fig:GMRT_low+high}), and we achieved an angular resolution of 21.55$''\times 17.3''$, comparable to the resolution of the LoTSS DR2 image (20$\arcsec \times 20\arcsec$) at 144 MHz in the low-resolution map. So, we used the LoTSS DR2 image at 20\arcsec and the uGMRT tapered image to create the spectral index map. Both the uGMRT and LoTSS DR2 images are convolved to a common resolution of $21.60\arcsec \times 21.60\arcsec$. Here, we did not include the structure before the break that contains pixels related to R1 and R2 (labelled in Fig. \ref{fig:LOFAR_high}) as it contains the mixing of plasma in low-resolution (around 20\arcsec). In this map, the spectral index of the extended tail varies from 0.3--0.8, with the flat spectral index on the hotspot-like structure near the outer edge of the tail. We used the Broadband Radio Astronomy ToolS \citep[BRATS;][]{harwood_2013MNRAS} to create the spectral index map. The images at two frequencies (144 MHz and 400 MHz) were aligned on a pixel-to-pixel basis, convolved to a common resolution, and then fitted using the \textsc{specindex} task in BRATS. Region selection and a 2$\sigma$ clipping were subsequently applied to generate the final spectral index map. The default weighted least-squares method was used for the fitting. Here, it should be noted that combining LoTSS DR2 and uGMRT images for spectral analysis can introduce systematic errors in the spectral index map due to differences in uv coverage. LoTSS DR2 has superior inner uv coverage (denser short-baseline sampling) compared to uGMRT, making it more sensitive to extended, low-surface-brightness emission. Because we cannot feasibly match the uv coverage of the LoTSS DR2 144 MHz map to that of uGMRT, some extended emission may be underestimated in the uGMRT data, leading to artificially flatter spectral indices in those regions.

\subsubsection{Radio luminosity measurements}
\label{sub:luminosity}
We have also calculated the radio luminosity ($R_{\textrm{L144}}$) for the various parts of the structure of J1007+3540 at 144 MHz and tabulated it in the 4th column of Table \ref{tab:radio_high}. For the calculation of radio luminosity, we have used the standard formula \citep[e.g.][]{Do09}
\begin{equation}	
    R_{\textrm{L144}}=4\pi{D_{L}}^{2}S_\mathrm{144}(1+z)^{\alpha-1},
\end{equation}
where $D_{L}$ is the luminosity distance to the source, $\alpha$ is the spectral index, $z$ is the redshift of J1007+3540, and $S_\mathrm{144}$ is the flux density at 144 MHz. As expected for an episodic radio galaxy, the radio luminosity of the inner lobes (south and north inner lobes) of J1007+3540 is lower, whereas the outer north lobe shows the highest radio luminosity among all parts \citep{Sc00a, Sc00b}. This is consistent with the outer north lobe exhibiting the highest jet kinetic power, indicating that the earlier outburst phase responsible for inflating this lobe was the most energetic episode of the source. The feedback or backflow structure and the southern extended tail exhibit moderately high jet kinetic powers, consistent with reprocessed plasma undergoing mixing, turbulence, and radiative losses. In contrast, the current inner jet shows the lowest jet kinetic power, suggesting that the active nucleus is presently in a weaker accretion state.

\subsubsection{Magnetic field and radiative age calculation}
\label{subsec:spec_age}
The magnetic field ($B_\mathrm{eq}$) calculations were done in accordance with the revised equipartition arguments provided by \citet[][see equation 3]{Be05}. For the calculation, the following assumptions were made: (a) the ratio of proton-to-electron number densities $K = 0$ (a relativistic pure electron--positron plasma), (b) the polarized emission originates from a regular magnetic field with all possible inclinations, and (c) the degree of polarization is assumed as 3 per cent.
The volume of the different regions was estimated using a cylindrical geometry. The obtained magnetic field strengths of the respective regions are tabulated in Table \ref{tab:radio_high}. The equipartition assumption is commonly employed to estimate magnetic field strengths in the absence of sufficient X-ray observations. However, studies have shown that the magnetic field strengths in the lobes of active FR II radio sources are typically lower by a factor of 2--3 compared to equipartition estimates \citep[e.g.;][]{Ha02, Cr05, Ka05, Mi07, Ine17, Tu18, Mi23}. Magnetic field strengths inferred from synchrotron modelling of radio emission and IC modelling of X-ray emission provide a quantitative measure of this discrepancy. This difference is often attributed to the presence of non-thermal particle populations or spatial variations in the particle energy distributions within the lobes, emphasizing the complex and inhomogeneous physical conditions present. However, \citet{Ko19} found that the magnetic field strengths in the inactive (outer) lobes of DDRGs were more consistent with those estimated using the equipartition method. This alignment in inactive lobes may indicate that, as energy injection ceases, the lobes evolve toward a state of energy balance over time.

\begin{figure}
\vbox{
\centerline{
\includegraphics[width=9cm,origin=c]{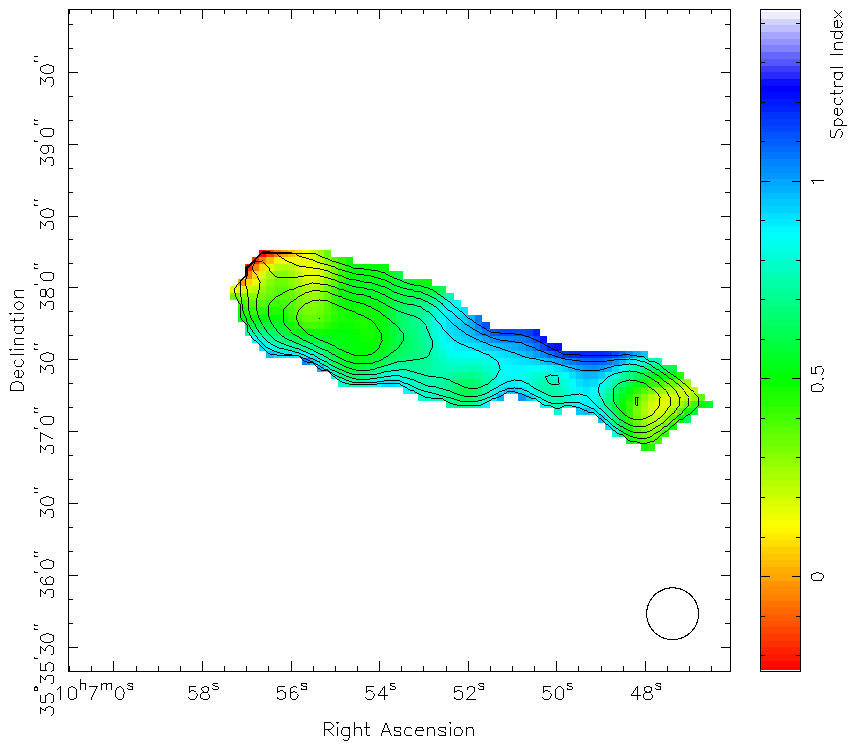}
}
}
\caption{Spectral index map of the extended tail of diffuse emission of J1007+3540 using LoTSS DR2 at 144 MHz and uGMRT at 400 MHz with $21.6\arcsec \times 21.6\arcsec$ resolution. The contour plot presents the LoTSS DR2 map of the extended tail of diffuse emission of \src{} at 144 MHz. The contour levels are at 3$\sigma \times (\sqrt{2})^{n}$, where $n$ = 0, 1, 2, 3,.....}
\label{fig:spec_age_low}
\end{figure}
Due to the availability of only two frequencies (144 and 400 MHz), we estimated the radiative age ($t_\mathrm{rad}$) of various parts of the structure of \src{} using Equation 2 of \citet{Br19}. As we have only two frequency data, it is not possible to directly measure the injection index by fitting the Jaffe--Perola (JP) \citep{Ja73} and Kardashev--Perola (KP) \citep{Ka62, Pa70} models. Therefore, we adopt a typical, physically motivated value of the spectral injection index $\alpha_\mathrm{inj} = 0.5$, for each part of the structure of J1007+3540 \citep{Pa70}. This value is predicted by shock acceleration theory \citep{Lo11} and has also been observed in regions of hot spots \citep[e.g.,][]{Ca91}. The injection spectral index is related to the power--law index of particles initial energy distribution \citep{Pa70}, and, in principle, its value can give us information about the properties of the jet--termination shock. In addition, \citet{Ko13} have shown that the values of the injection index (the low-frequency power--law index of synchrotron emission) are similar in different episodes of jet activity. The $t_\mathrm{rad}$ are provided in the seventh column of Table \ref{tab:radio_high}. The $B_\mathrm{eq}$ used in the $t_\mathrm{rad}$ estimation, as well as the corresponding break frequencies ($\nu_{\mathrm{br}}$) (as per the assumptions in \citet{Br19}) for J1007+3540, are listed in the fifth and sixth columns of Table \ref{tab:radio_high}. As noted, the $t_\mathrm{rad}$ of the inner lobes is $\sim$140 Myr, associated with $\nu_{\mathrm{br}}$ $\sim$500 MHz. In contrast, the outer north lobe and the feedback emission are significantly older, with $t_\mathrm{rad}$ of $\sim$240 Myr and $\sim$260 Myr, and low $\nu_{\mathrm{br}}$ of 155 MHz and 138 MHz, respectively. The large extended tail of diffuse emission of \src{} shows $t_\mathrm{rad}$ of 100 Myr and a $\nu_{\mathrm{br}}$ of 457 MHz. According to the JP model \citep{Ja73}, the $t_\mathrm{rad}$ is inversely proportional to the square root of $\nu_{\mathrm{br}}$ ($t_\mathrm{rad} \propto \nu_{\textrm{br}}^{-1/2}$). Therefore, as $\nu_{\mathrm{br}}$ decreases, $t_\mathrm{rad}$ increases (indicating the older plasma) and vice versa, which is consistent with the trend observed across the presented source J1007+3540, except for the plasma in the large extended tail of diffuse emission.

\begin{figure*}
\vbox{
\centerline{
\includegraphics[width=17cm,origin=c]{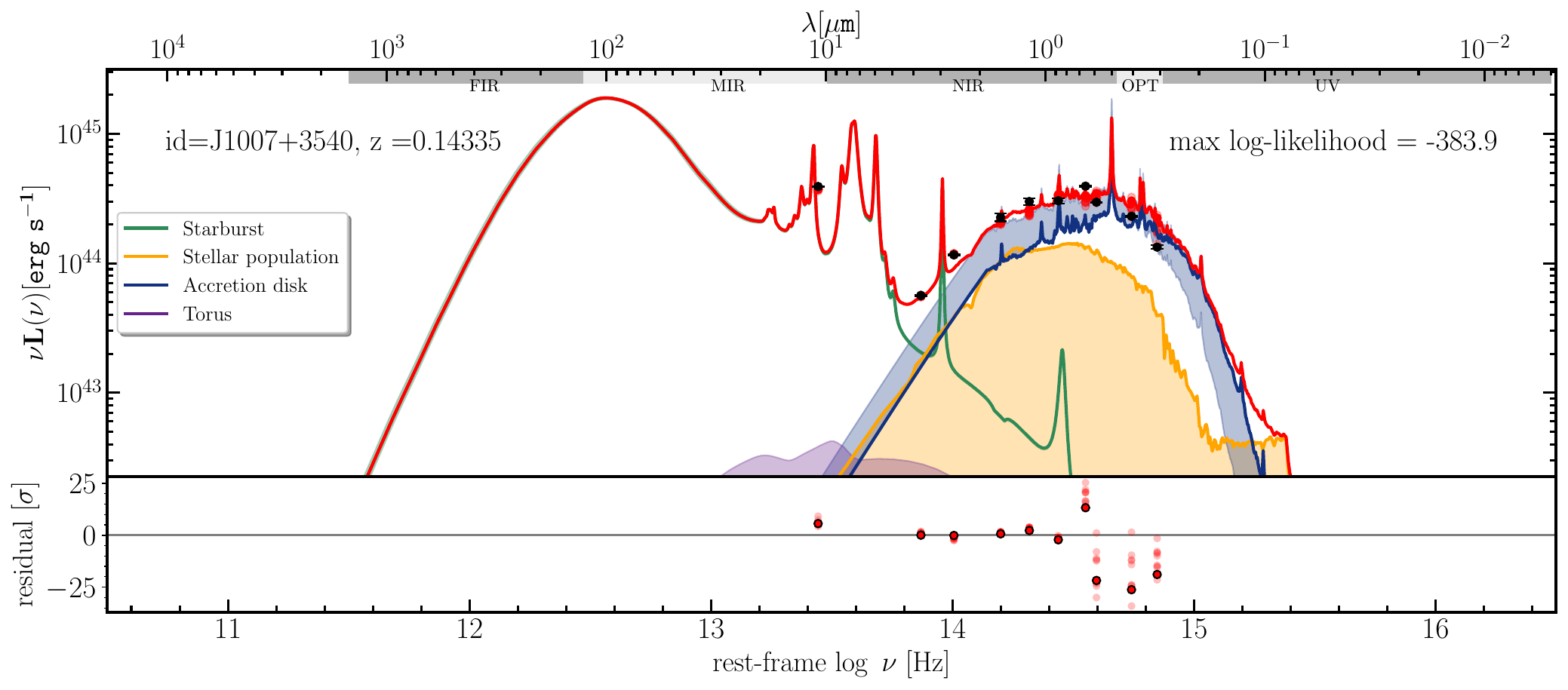}
}
}
\caption{Multi-component spectral energy distribution (SED) fitting of the host galaxy (MaxBCG J151.77665+35.67813) of GRG J1007+3540 using the \texttt{AGNfitter} code \citep{Ca16, Ma24} covering the mid-infrared (MIR), near-infrared (NIR), and optical domains.}
\label{fig:SED}
\end{figure*}

\begin{table}
\centering
\caption{SED-derived physical parameters for the central active SMBH (without including contributions from the extended jet or lobe structures) for J1007+3540 with the respective 84th percentiles.}
\label{tab:agnfitter_output}
\begin{tabular}{llll}
\hline
Parameter & Symbol & Value & Unit  \\
\hline
Stellar mass & $\log_{10}(M_\star/M_\odot)$ & 11.0 & dex  \\
Stellar population age & Age$_{\rm gal}$ & $12.4^{+1.4}_{-2.3}$ & Gyr \\
Star formation timescale & $\tau$ & $11^{+4}_{-6}$ & Gyr  \\
Star formation rate (optical) & $\log_{10}(\mathrm{SFR}_{\rm opt})$ & 0.84 & $M_\odot~ \mathrm{yr}^{-1}$  \\
Star formation rate (infrared) & SFR$_{\rm IR}$ & 106 & $M_\odot~ \mathrm{yr}^{-1}$  \\
Dust extinction (galaxy) & $E(B-V)_{\rm gal}$ & 0.64$^{+0.24}_{-0.21}$ & mag  \\
Dust extinction & $E(B-V)_{\rm bbb}$ & 0.89$^{+0.06}_{-0.18}$ & mag  \\
Dust temperature & $T_{\rm dust}$ & 30.9$^{+0.5}_{-0.7}$ & K \\
Bolometric luminosity & $\log_{10}(L_{\rm AGN})$ & 44.8 & erg~s$^{-1}$  \\
\hline
\end{tabular}
\end{table}

\subsection{Host galaxy properties and AGN mode}
\label{subsec:SED_and_AGN}
We presented the SED fitting of MaxBCG J151.77665+35.67813 in
Fig. \ref{fig:SED}, where the black points represent the broad-band photometric
data covering the mid-infrared (MIR), near-infrared (NIR), and
optical domains. The derived SED fitting parameters are tabulated in Table \ref{tab:agnfitter_output}. The derived stellar population age is approximately $12.4^{+1.4}_{-2.3}$ Gyr, suggesting that the host galaxy formed its stars relatively early and is now an evolved system. The inferred stellar mass is $\log_{10}(M_\star/M_\odot) = 11.0$. The star formation rate estimated from the optical data is modest, at $\mathrm{SFR}_{\mathrm{opt}} \approx 0.84~M_\odot~ \mathrm{yr}^{-1}$, while the infrared-based star formation rate is significantly higher, with $\mathrm{SFR}_{\mathrm{IR}} \approx 106~M_\odot~ \mathrm{yr}^{-1}$. This large discrepancy highlights the presence of heavy dust obscuration, with most star-forming activity being hidden from optical view and revealed only in the IR. The star formation timescale for the host galaxy is estimated as $11^{+4}_{-6}$ Gyr.

Dust extinction is found to be substantial, with a colour excess of $E(B-V)_{\mathrm{gal}} = 0.64$ for the stellar component and $E(B-V)_{\mathrm{bbb}} = 0.89$ for the AGN accretion disk. The dust temperature is measured to be $T_{\mathrm{dust}} = 30.9$ K, consistent with active star-forming galaxies. The bolometric luminosity of AGN, integrated over the 0.4--12 $\mu$m range, is $\log_{10}(L_{\mathrm{AGN}}~[\mathrm{erg ~s^{-1}}]) = 44.8$, corresponding to $L_{\mathrm{AGN}} \approx 6.0 \times 10^{44}~\mathrm{erg ~s^{-1}}$. This high luminosity, together with the presence of strong torus emission, and the presence of the ``Big Blue Bump'' suggests that the AGN is currently in a radiatively efficient (RE) AGN, though it is significantly obscured.

For the source J1007+3540, the jet position angle, derived from the core and lobe coordinates, is measured to be approximately $29.7^\circ$. The host galaxy, identified as PGC~2068293, has a major-axis PA of $72.1^\circ$, which places the minor axis at $162.1^\circ$. The tentative angle between the jet and the minor axis in projection is $\Delta\theta = |29.7^\circ - 162.1^\circ| = 132.4^\circ$. This indicates that the jet is moderately misaligned with the stellar structure of the host galaxy, neither fully perpendicular nor parallel to it. However, this measurement is based on 2D position angles on the sky, and the true 3D orientation of the jet relative to the host galaxy cannot be determined from these data alone. 

We used publicly available data obtained from the Large Sky Area Multi-Object Fiber Spectroscopic Telescope \citep[LAMOST;][]{Cu12} to estimate the mass of the central supermassive black hole of the host galaxy MaxBCG J151.77665+35.67813. These spectroscopic data were previously used by \citet{Na20} to derive the stellar velocity dispersion ($\sigma^{\ast}$) for a large subsample of LAMOST galaxies. From the LAMOST Low Resolution Search (LRS) Catalogue of Stellar Population Synthesis of Galaxies \footnote{https://www.lamost.org/dr10/v2.0/table/galaxy}, we obtained a value of $\sigma^{\ast}=336\pm12$ km s$^{-1}$ for our host galaxy (in this catalogue, it is an object with ID 400905194 at RA$=$151.776642\degr; Dec $=+$35.678131\degr).

We estimated the black hole mass as $M_\mathrm{BH}=3.0\pm0.6\times10^9 M_{\odot}$ using the $M_\mathrm{BH}-{\sigma^{\ast}}$ relation \citep{Fe00, Ge00} (see equation \ref{eq:MBH}) and adopted the constants derived in \citet{Kormendy13}. 
\begin{equation}
\label{eq:MBH}
\frac{M_\mathrm{BH}}{10^{9}~M_{\odot}} = 
\left(0.310^{+0.037}_{-0.033}\right)
\left(\frac{\sigma^{\ast}}{200~\mathrm{km~s^{-1}}}\right)^{4.38 \pm 0.29}
\end{equation}
where $M_{\odot}$ is the stellar mass and $\sigma^{\ast}$ is the velocity dispersion. The estimated $M_\mathrm{BH}$ for the host galaxy of J1007+3540 is slightly higher than, the mean $M_\mathrm{BH}$ for a GRG sample obtained by \citet{Da20b}: $M_\mathrm{BH}=1.04\times10^9 M_{\odot}$ and by \citet{Oe22}: $M_\mathrm{BH}=1.5\times10^9 M_{\odot}$. Using the blackhole mass, we have estimated the Eddington magnetic field \citep{Be10, Dal11} as $B_{\mathrm{Edd}} = 1.15 \times 10^{4}~\mathrm{G}$ (where
$B_{\mathrm{Edd}} = 6.1 \times 10^{4} 
\left( \frac{M_{\mathrm{BH}}}{10^{8} M_{\odot}} \right)^{-1/2}~\mathrm{G} $).
According to the Blandford--Znajek (B-Z) mechanism \citep{Bl77}, 
the jet power depends on the black hole spin ($a$), magnetic field ($B$), 
and black hole mass ($M_{\mathrm{BH}}$) as
$
Q_{\mathrm{jet}} \propto a^{2} B^{2} M_{\mathrm{BH}}^{2} $.
Adopting a proportionality constant $s = \sqrt{0.5}$, 
the spin parameter can be expressed as
$
a = s \times 
\sqrt{\frac{Q_{\mathrm{jet}}}{B_{\mathrm{Edd}}^{2} M_{\mathrm{BH}}^{2}}}.$ The derived spin parameters range from $a \approx 0.002$ to $0.005$ across the different morphological parts of J1007+3540, indicating an overall low-spin state for the central black hole. Such low values are typical of massive, slowly rotating black holes accreting at low Eddington ratios.

\subsection{X-ray emission near J1007+3540}
\label{subsec:X-ray}
The thermal gas in galaxy clusters is most directly traced by the diffuse X-ray emission forming the cluster halo. However, we found no deep observation for this cluster in any of the X-ray telescope data archives. The only available X-ray map is the "Spectrum-Roentgen-Gamma" (SRG)/extended ROentgen Survey with an Imaging Telescope Array (eROSITA) all-sky survey (eRASS1) map. Thus, to study the X-ray emission near J1007+3540, we used the eRASS1 map of the source \citep{Sunyaev2021, Predehl21}.
 An eRASS1 image of J1007+3540 is overlayed with a low-resolution uGMRT image (see Fig. \ref{fig:eROSITA_map}) in black contour. We smoothed the eRASS1 image at a resolution of 30 arcsec. The eRASS1 image reveals extended X-ray emission (presented in the cyan contour in Fig. \ref{fig:eROSITA_map}). This overlay highlights the spatial alignment of diffuse X-ray emission with the radio morphology of the source J1007+3540, particularly around the central region. The concentration of X-ray emission near the cluster centre aligns with the galaxy distribution (see Fig. \ref{fig:cluster_galaxies_plot}), suggesting that these galaxies are possibly interacting with the ICM and may be contributing to the discontinuity of the extended tail diffuse emission of J1007+3540. We noted the X-ray luminosity for the X-ray emission from the eRASS1 map at 0.2--2.3 keV \citep{Bu24} as $L_{X}=3.2\pm0.3\times10^{43}$~erg~s$^{-1}$ (see the third column of Table \ref{Tab:X-ray}). We estimated the temperature of the plasma ($T_X$) as 1.80 keV ($2.1\times10^{7}~\rm{K}$) using the relation $T_{X}=2.34\;L_{44}^{1/2}\;h_{50}$ $\mathrm{keV}$ \citep{Bo00} where $L_{44}$ is the X-ray luminosity, expressed in units of $10^{44}$~erg~s$^{-1}$, within the 0.2--2.3 keV energy band and $h_{50}$ is the Hubble constant in units of 50 km s$^{-1}$ Mpc$^{-1}$.

\section{Discussions}
\label{sec:discuss}
In this section, we discuss the various unusual parts of the structure of J1007+3540 and the possible scenarios for their formation.
\subsection{Recurrent/episodic jet activity in J1007+3540}
\label{subsec:recurrent}
An episodic RG refers to a source that has experienced multiple phases of jet activity, separated by quiescent periods where the jets shut down before reactivating \citep{Cl91, Cl92, Pe94, Sc00a, Sc00b}. This results in both older (relic) and newer (active) radio lobes. The outer lobes of an episodic radio galaxy are typically diffuse and aged, reflecting an earlier phase of activity, while the inner lobes are brighter, more compact, and represent a more recent phase. The outer lobes often have a steeper radio spectrum due to synchrotron and IC CMB ageing \citep[e.g.][]{Hardcastle2004, Mingo}, while the inner jets exhibit a flatter spectrum, with active hotspots at their ends. For the source J1007+3540, detailed in Section \ref{sub:radio_pro} and tabulated radio properties in Table \ref{tab:radio_high}, the outer north lobe is steeper ($\alpha$ range of 1.9--2.5) than the inner lobes ($\alpha$ range of 0.85--1.1), which is consistent with the characteristics of a typical episodic RG. Notably, the outer north lobe is much steeper than the outer south extended tail of diffuse emission ($\alpha$ range of 0.3--0.8; see Fig. \ref{fig:spec_age_low}). 
\begin{figure}
\vbox{
\centerline{
\includegraphics[width=9cm,origin=c]{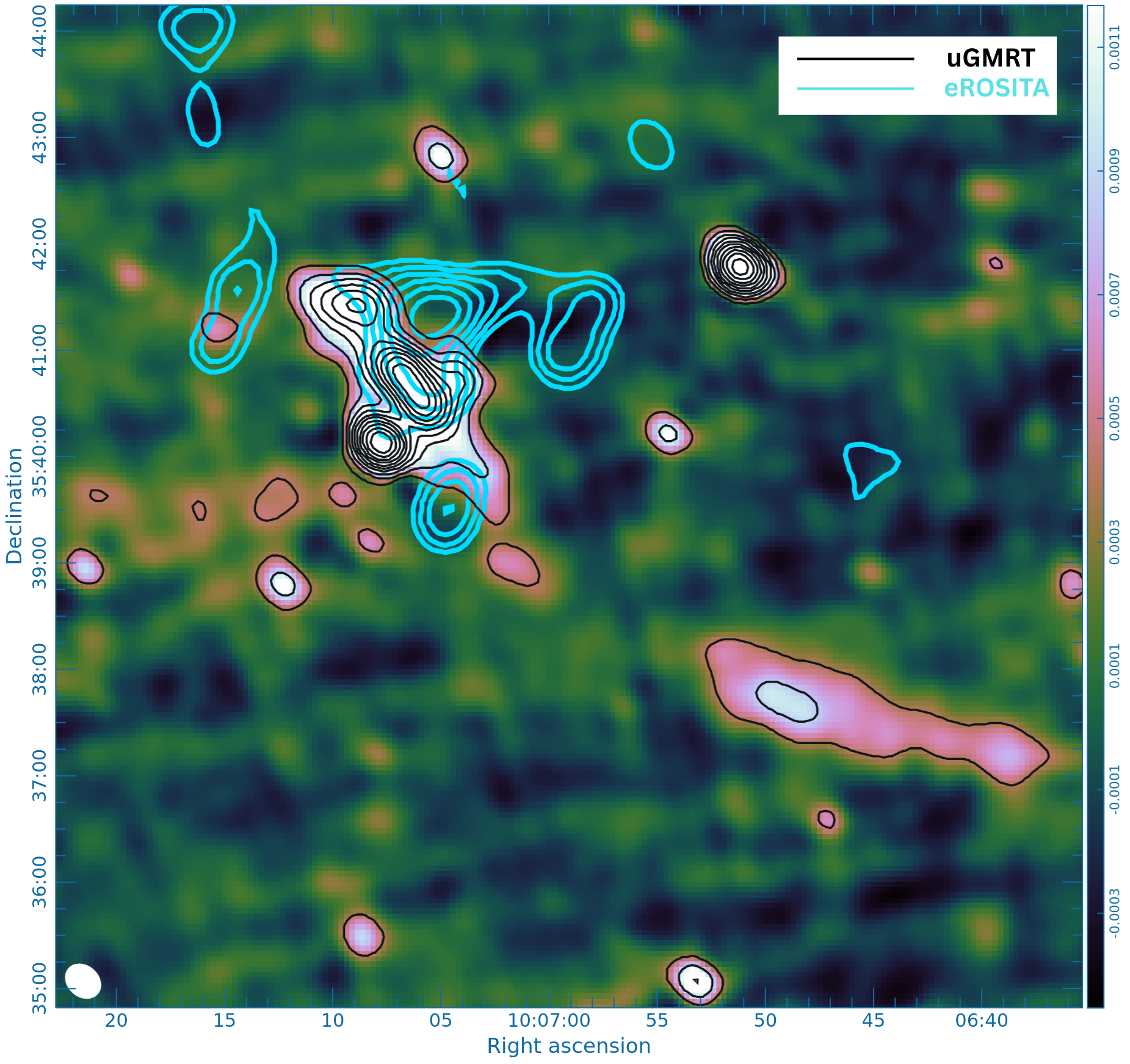}
}
}
\caption{eROSITA image of \src{} (cyan contour) overlaid with uGMRT image at 400 MHz (background image and black contour). For the uGMRT image at 400 MHz, contour levels are at 0.09, 0.19, 0.32, 0.47, 0.67, 0.91, 1.22, 1.60, 2.07, 2.67, 3.42, 4.35, 5.51, 6.97, 8.78 mJy beam$^{-1}$. The colour scale represents the intensity of the source in Jy~beam$^{-1}$ at the 400 MHz uGMRT image.}
\label{fig:eROSITA_map}
\end{figure}
The radiative age analysis of \src{} reveals distinct variations in synchrotron ageing across its parts. The inner lobes are younger with radiative ages of $\sim$140 Myr. The outer north lobe appears significantly older with a radiative age of 240--260 Myr \citep{Jamrozy2007}. This stark difference ($\sim$100 Myr) \citep{Ma10} suggests that the outer north lobe is a relic of a previous jet activity cycle, while the inner lobes result from a more recent phase of AGN jet activity \citep{Jamrozy2007, Ko13, Ko19}. Under the assumption of passive ageing, the spectral age of the outer north lobe indicates it has not undergone significant recent re-energization; however, we cannot rule out the possibility of mild re-acceleration that could slightly modify the inferred age difference relative to the inner lobes. Such age differences highlight the temporal separation between distinct phases of AGN activity. Interestingly, in typical episodic RGs, inner lobes are expected to be much younger \citep{Jamrozy2007, Ko13, Ko19} than what is observed here for J1007+3540. This discrepancy raises the possibility that the very steep spectrum backflow emission, characteristic of aged plasma, may be contaminating or blending with the spectra of the inner lobes in \src{}.

The optical properties of the host galaxy also reinforce this episodic picture. The SED fitting of MaxBCG J151.77665+35.67813 reveals an old stellar population ($\sim$12 Gyr) and a massive system with $\log_{10}(M_\star/M_\odot) \approx 11.0$ \citep{Br03, Conroy2009}. 
Despite its evolved stellar content, the host shows evidence of substantial dust [$E(B-V)_{\rm gal} \sim 0.6$] and a high infrared-based star formation rate ($\sim$100$~M_\odot~\mathrm{yr}^{-1}$) \citep{Kennicutt1998, Kennicutt2012}. Such dusty, gas-rich conditions provide a potential reservoir for intermittent fueling of the central black hole \citep{Fabian2012, He14}. In this context, the recurrent jet activity in J1007+3540 may be explained by cycles of gas inflow into the nucleus, which temporarily shuts down and later retriggers the AGN \citep{Sa09,Ko13}. The long star formation timescale ($\tau \sim 11$ Gyr; see Table \ref{tab:agnfitter_output}) further suggests that the stellar population has evolved over cosmic time with episodic bursts of jet activity sustained by external or recycled gas \citep{Thomas2005, Ciotti2007}. 
This supports a scenario where the interplay between the dust-obscured star-forming activity in the host, the surrounding cluster medium, and feedback-driven gas dynamics collectively regulates the recurrent AGN behavior observed in J1007+3540 \citep{McNamara2007, Fabian2012}.

\subsection{Distorted backflow signature in the northern lobe}
\label{subsec:relic} 
The outer north lobe of J1007+3540 shows clear evidence of relic emission combined with a distorted backflow signature directed toward the southeast (see Fig. \ref{fig:LOFAR_high} and the left panel of Fig. \ref{fig:GMRT_low+high}). The northern lobe with distorted backflow emission is steeper ($\alpha$ range of 1.9--2.5) and more aged (240--260 Myr) than the other parts of the structure of J1007+3540, consistent with passively ageing plasma that has not been re-energized. Such backflowing plasma is often observed when radio jets encounter a dense surrounding medium, causing the lobe material to be deflected and redirected away from the jet axis \citep{Leahy1991, Mizuta2004, Cielo2014}. The compressed appearance of the outer north lobe supports this interpretation: plasma appears to flow back toward the centre after interacting with the ICM, a process commonly seen in cluster--embedded radio galaxies \citep[e.g.][]{Perucho2019, Hardcastle2020}. As this material approaches the host galaxy, it meets the newly formed inner episodic jet, resulting in distortions of the overall jet--lobe morphology.
\begin{table}
\begin{center}
\caption{Properties of X-ray emission measured within $\rm R_{500}$ (see Fig. \ref{fig:eROSITA_map}) obtained from the eROSITA All-Sky Survey (eRASS1) map at 0.2--2.3 keV \citep{Bu24}.}
\begin{tabular}{cccc}
\hline
 Count rate & Flux  &  $\rm L_X$ & $\rm T_X$ \\  
  (ct~s$^{-1}$)  &(mWm$^{-2}$)& (erg s$^{-1}$ ) & (keV)\\  
  (1)  & (2) & (3) & (4)  \\ \hline 
 0.61 & 57.5 & 3.20 $\times$ $\rm 10^{43}$ & 1.80\\
\hline
    \\
\end{tabular}
\\
NOTE. (1) Count rate observed within $\rm R_{500}$ (2) Flux within $\rm R_{500}$ (3) L$_{\rm X}$: X-ray luminosity within $\rm R_{500}$ and (4) T$_{\rm X}$: Temperature in the X-ray band within $\rm R_{500}$. 
\label{Tab:X-ray}
\end{center}
\end{table}
We used the eROSITA map, presented in Fig. \ref{fig:eROSITA_map}, to study the X-ray emission near the WHL 100706.4+354041 cluster. The eROSITA map detected low-level diffuse X-ray emission near the cluster centre, with a luminosity of $L_{X}=3.20\pm0.3\times10^{43}$~erg~s$^{-1}$ and a correspondingly low gas temperature of $\sim$1.8 keV. Due to non-uniform distribution and point source contamination in the map, the eROSITA low-resolution map is not suitable for detecting any shock. Therefore, to study the thermal emission of the gas and its interaction with the surrounding medium in greater detail, high-resolution and sensitive X-ray observations with the \emph{Chandra} X-ray Observatory are required. The sub-arcsecond spatial resolution of Chandra is particularly well-suited for resolving small-scale structures such as cavities, shocks, or cold fronts \citep[e.g.][]{McNamara2007, Fabian2012}. Alternatively, \emph{XMM--Newton} can provide complementary insights due to its higher sensitivity to diffuse emission, especially useful for mapping the extended thermal gas \citep{Snowden2008}.

\subsection{Re-acceleration and jet--ICM interaction in J1007+3540}
\label{subsec:extended_tail}
In addition to the relic northern lobe, J1007+3540 exhibits a remarkable extended tail of diffuse emission on its southwestern side, with a distinct break in morphology. High-sensitivity uGMRT tapered image (with low-resolution; $21.55'' \times 17.3''$, RMS$\sim$30 $\mu$Jy~beam$^{-1}$) at 400 MHz (see right panel of Fig. \ref{fig:GMRT_low+high}; bright radio components) reveals a hotspot-like structure at the outer edge of the extended tail. This hotspot-like structure in the outer edge of the extended tail of J1007+3540, also seen in the spectral index map (see Fig. \ref{fig:spec_age_low}), is similar to a typical FR II RG, although no associated optical counterpart or core is detected at its centre. At lower frequency (with high resolution; $6\arcsec \times 6\arcsec$), the LoTSS DR2 map at 144 MHz resolves this region as faint diffuse emission without any hotspots (see Fig. \ref{fig:LOFAR_high}). A compressed lobe is seen just before the break in the extended tail (see Figs. \ref{fig:LOFAR_high} and \ref{fig:LOFAR_low}). This compressed lobe suggests that the jet plasma may be disrupted or deflected by interaction with the ICM, leading to the elongated, diffuse morphology observed beyond the break \citep{Hardcastle2004}. 

A striking feature of this southwestern extended tail is its spectral and radiative age behaviour. The extended tail beyond the break shows a lower spectral index ($0.3-0.8$; see Fig. \ref{fig:spec_age_low}) than expected, and its estimated radiative age ($\sim$100 Myr) is surprisingly younger than that of the inner episodic lobes ($\sim$140 Myr). We hypothesize that the diffuse emission may interact with the surrounding cluster medium, potentially flattening the spectral index. This counterintuitive result provides strong evidence for re-acceleration mechanisms acting on an older electron population in this region. While the southwestern tail provides the clearest evidence for re-acceleration, it is possible that similar processes are occurring in other diffuse emission (lobes) of the source, but remain undetected due to lower surface brightness, projection effects, or insufficient spectral coverage. Possible processes include turbulence induced by jet--ICM interactions, weak shocks propagating through the plasma, or entrainment of cluster gas, all of which can rejuvenate the electron population and reduce the apparent synchrotron age \citep{En01, Markevitch2005, vanWeeren2019}. We overplotted the positions of optical sources within the R$_{500}$ radius with the LoTSS DR2 map at 144 ($6\arcsec \times 6\arcsec$; see Fig. \ref{fig:cluster_galaxies_plot}), and the location of the extended tail relative to the cluster centre further supports this view. It is found that some part of the diffuse extended tail emission appears to extend toward or even beyond the $R_{500}$ radius (see Fig. \ref{fig:cluster_galaxies_plot}), indicating that the jets are propagating into the outskirts of the cluster medium. The combination of a break in morphology, spectral flattening, and anomalously young radiative age points to a complex history of jet--ICM interaction. This makes the diffuse tail more fascinating to study re-acceleration physics in cluster-surrounded giant radio galaxies.

\section{Summary and conclusions}
\label{sec:conc}
 In this paper, we presented a detailed multi-waveband analysis of a 1.45 Mpc GRG J1007+3540. The source is hosted by MaxBCG J151.77665+35.67813, situated within the WHL J100706.4+354041 galaxy cluster. This GRG is identified in the LoTSS DR2 map at 144 MHz. To explore the radio morphology and spectral characteristics in detail, we performed dedicated observations of the source using the uGMRT. We found strong evidence of episodic jet activity in J1007+3540. The inner lobes of J1007+3540 exhibit a spectral index range of 0.85--1.1, while the outer north lobe is characterized by ultra-steep spectrum ($\alpha$ range of 1.9--2.5), indicative of relic plasma. Radiative age estimates support this interpretation, suggesting that the inner lobes are significantly younger, with ages of $\sim$140 Myr, compared to the outer north lobe and backflow plasma, which have radiative ages of $\sim$240 Myr and $\sim$260 Myr, respectively. This temporal offset of more than 100 Myr between the inner and outer parts of the structure of J1007+3540 indicates multiple episodes of jet launching. Additionally, we detect an extended tail of diffuse emission in the southern direction, followed by a morphological break. The region before the break shows an ultra-steep spectral index ($\alpha \sim 3.6$), while beyond the break extended tail displays a flat spectral index ($\alpha$ range of 0.3--0.8) and a younger radiative age ($\sim$100 Myr) compared to the other parts of J1007+3540. This suggests that the extended tail after the break may be experiencing re-acceleration processes, possibly due to interaction with the ICM.

To further characterize the physical conditions of the host galaxy and AGN, we performed SED modelling. The SED analysis provides a complementary perspective, showing that the host galaxy is an evolved elliptical system with high levels of dust extinction. The radiatively efficient AGN, together with its high infrared star formation rate, suggests an obscured yet active central engine. 
These conditions are consistent with merger-driven fueling or a rejuvenated accretion episode, potentially linked to the observed jet restarting. Importantly, the SED modelling supports the idea that both AGN activity and star formation are coexisting in a dynamically evolving environment, shaped by both internal processes and external environmental factors.

Despite the lack of deep X-ray imaging and additional intermediate-frequency radio data, our study reveals valuable insights about GRG  J1007+3540 and its interplay with the surrounding cluster medium. The morphological asymmetries, jet bending, backflow plasma, episodic AGN activity, a large tail of diffuse emission with a morphological break, spectral gradients, and radiative age distribution collectively point toward a complex history of jet--ICM interaction and a possible re-acceleration process occurring in the surroundings of J1007+3540. These findings emphasize the importance of future deep, multiwavelength studies, including X-ray, optical, and multiwavelength radio observations, to further probe the dynamics and energetics of this interesting source.

\section*{Acknowledgements}
We thank the anonymous reviewer for helpful suggestions.
S. Kumari gratefully acknowledges the Department of Science \& Technology, Government of India, for financial support, vide reference no. DST/WISE-PhD/PM/2023/3 (G) under the `WISE Fellowship for Ph.D.' program to carry out this work. S. Paul acknowledges the Manipal Centre for Natural Sciences, Centre of Excellence, and Manipal Academy of Higher Education (MAHE) for facilities and support. This paper makes use of data of LOw-Frequency ARray (LOFAR) Two-metre Sky Survey second data release (LoTSS DR2) available at \href{https://lofar-surveys.org/}{https://lofar-surveys.org/}. LOFAR data products were provided by the LOFAR Surveys Key Science project (LSKSP; \href{ https://lofar-surveys.org/}{ https://lofar-surveys.org/}) and were derived from observations with the International LOFAR Telescope (ILT). LOFAR \citep{va13} is the Low-Frequency Array designed and constructed by ASTRON. It has observing, data processing, and data storage facilities in several countries, which are owned by various parties (each with their funding sources), and which are collectively operated by the ILT foundation under a joint scientific policy. The efforts of the LSKSP have benefited from funding from the European Research Council, NOVA, NWO, CNRS-INSU, the SURF Co-operative, the UK Science and Technology Funding Council (STFC) and the Jülich Supercomputing Centre. We thank the staff of the GMRT who made these observations possible. GMRT is run by the National Centre for Radio Astrophysics of the Tata Institute of Fundamental Research. We also acknowledge the use of the LAMOST survey. Guoshoujing Telescope (the Large Sky Area Multi-Object Fiber Spectroscopic Telescope, LAMOST) is a National Major Scientific Project built by the Chinese Academy of Sciences. Funding for the project has been provided by the National Development and Reform Commission. LAMOST is operated and managed by the National Astronomical Observatories, Chinese Academy of Sciences. 

\section*{Data Availability}
This paper makes use of GMRT data available at \href{https://naps.ncra.tifr.res.in/goa/data/search}{https://naps.ncra.tifr.res.in/goa/data/search}. 
This publication also uses Pan-STARRS1 data available at \href{http://ps1images.stsci.edu/cgi-bin/ps1cutouts}{http://ps1images.stsci.edu/cgi-bin/ps1cutouts}. The LOFAR data availability is described under `Acknowledgements'.











\bsp	
\label{lastpage}
\end{document}